\documentclass[prd,reprint,prd,twocolumn,superscriptaddress,nofootinbib]{revtex4-2} 

\usepackage{graphicx,braket}
\usepackage{multirow}
\usepackage{amsmath,amssymb,amsfonts}
\usepackage{amsthm}
\usepackage{mathrsfs}
\usepackage[title]{appendix}
\usepackage{xcolor}
\usepackage{textcomp}
\usepackage{mathtools}
\usepackage{booktabs}
\usepackage{algorithm}
\usepackage{algorithmicx}
\usepackage{algpseudocode}
\usepackage{listings}
\AtBeginDocument{
  \setlength{\abovedisplayskip}{5pt plus 1pt minus 1pt}
  \setlength{\belowdisplayskip}{5pt plus 1pt minus 1pt}
  \setlength{\abovedisplayshortskip}{2pt plus 1pt minus 1pt}
  \setlength{\belowdisplayshortskip}{2pt plus 1pt minus 1pt}
}
\allowdisplaybreaks

\newtheorem{theorem}{Theorem}

\newtheorem{example}{Example}

\newtheorem{definition}{Definition}

\begin{document}


\title{Wave packets asymptotics in $\kappa$-minkowski space}

\author{Adith A}
 \affiliation{Department of Physics, National institute of technology Calicut, Calicut, 673601, India}
 \email{adithammath@gmail.com}
\author{A. Bhagwat}
 \email{ameeya@cbs.ac.in}
\affiliation{
School of Physical Sciences, UM-DAE Centre for Excellence in Basic Sciences, University of Mumbai, Vidyanagari campus, Mumbai 400098, India
}
\affiliation{Centre for Excellence in Theoretical and Computational Sciences, University of Mumbai, Vidyanagari Campus, Mumbai 400098, India}

\date{\today}

\begin{abstract}
In this paper, we study the nature of wave packets and their decay behavior in the $\kappa$-Minkowski
space. We employ techniques from singularity theory throughout our analysis. Specifically, we
use methods from the theory of Lagrangian singularities and caustics to study the topology of the
energy-level manifold. The Newton polyhedron method from the theory of oscillatory
integrals is applied to analyze the decay behavior of wave packets. We find that for a generic Hamiltonian in
the $\kappa$-Minkowski space, the associated wave packets exhibit different decay rates depending on the
control parameter of the system. When the control parameter takes values on the cusp, the system
exhibits decay of order $t^{-1/4}$; otherwise, it exhibits decay of order $t^{-1/2}$. This distinctive behavior of wave packets in the $\kappa$-Minkowski makes it fundamentally different from commutative spacetime.
\end{abstract} 

\maketitle

\section{Introduction}\label{sec1}
Noncommutative geometry, an area of mathematics, studies spaces that do not have the properties that characterize them as `classical' spaces (like measure spaces,locally compact spaces or smooth spaces)\cite{connes1982survey,connes1980calgebres}. For example, the spaces which can be obtained by quotienting smooth manifolds with ergodic groups fall into the research interest of noncommutative geometry. The quotient spaces that are studied may not even classify as Hausdorff spaces, let alone be smooth. The first rigorous mathematical study of noncommutative geometry was carried out by Alain Connes \cite{connes1995noncommutative,Connes2}. His functional analytic approach, which is based on the notion of $C^*$ algebra, serves as the foundation of the subject. The hallmark theorem was an extension of the  Gel’fand-Naimark theorem
\footnote{Every commutative $C^*$-algebra is $C*$-isomorphic to $C(K)$ for some compact Hausdorff space K. Specifically for a Commutative $C^*$ algebra $\mathcal{A}$ the Gelfand transform is a $C^*$ isomorphism from $\mathcal{A}$ onto $C(\mathcal{M}_{\mathcal{A}})$\cite{zhu1993introduction}} .

Alain Connes extended this theorem so that a noncommutative space can be seen as a noncommutative $C^*$ algebra and vise versa, thus establishing a duality between them. In his study he was able to represent a noncommutative spacetime by a spectral triple consisting of noncommutative algebra, a Hilbert space, and a special operator called the Dirac operator. Major research has been done to construct spectral triples for various hypothesized noncommutative spaces modeling spacetime. However, the idea of using noncommutative spaces in physics was put forward back in 1930 by Heisenberg even before mathematicians studied them \cite{PauliBriefwechselVol2,PauliBriefwechselVol3}(Heisenberg did not publish these but communicated his ideas with Pauli who then introduced it to Oppenheimmer; Oppenheimmer passed on the puzzle to his student Synder, who finally formalized the subject\cite{Synder}). He argued that just as we have a nontrivial commutation relation between position and momentum making the phase space noncommutative, we can also make the commutator between position coordinates non-zero. He believed that this might solve some of the divergence issues in quantum field theory by introducing a natural uncertainty in $\Delta x$, but later it was realised that introducing noncommutativity will not solve the divergence issues but rather will introduce UV-IR mixing, therefore, this was replaced by renormalization schemes \cite{Synder}. The interest in noncommutative geometry was revived because of connections to quantum gravity theories. Noncommutative gauge theory on Moyal spacetime has been shown to arise as a limit of certain models of string theory \cite{NathanSeiberg_1999}.

There are different models of noncommutative spaces that are being studied extensively, an example of which is the Moyal spacetime\cite{GROENEWOLD1946405} which assumes the commutator algebra of spacetime coordinates to be $[X_\mu ,X_\nu]=i\theta_{\mu \nu}$. Even though it is a successful model, Moyal spacetime suffers from the issue of Lorentz covariance\cite{AMELINOCAMELIA2001255,DSRkeyresults}, as a result alternative models have been suggested, one of them being a Lie algebra type model called $\kappa$-deformation \cite{Meljanac2006,kappaintro,MAJID1994348} whose basic commutator algebra satisfies $[X_\mu ,X_{\nu}]=i C^{\lambda}_{\mu \nu}X_\lambda$. This model of spacetime arises naturally in a model of relativity called Double special relativity (DSR)\cite{AMELINOCAMELIA2001255,DSRkeyresults} which has both $c$ and $\hbar$ as fundamental observer independent quantities.

Singularity theory\cite{Arnold1998} is a branch of mathematics that deals with critical points of functions and their behavior near those points, and it has applications in various branches of physics such as dynamical systems and catastrophes\cite{Arnold1998}. It is used especially in conditions where the properties of the system change drastically when control parameters are changed in a continuous manner. In the theory of singularities, the area of utmost fascination is that of the theory of Lagrangian singularities \cite{Arnold2012}. This branch studies the critical points arising when we project (henceforth called Lagrangian projections) a Lagrangian submanifold of a phase space onto the configuration space. Take as an example the one dimensional harmonic oscillator. The level sets of this system (i.e., the curve $\mathcal{H}=E$) form a Lagrangian submanifold and when we project this down onto the configuration space we get two fold singularities which are exactly the turning points of the system where $\partial \mathcal{H}/\partial p=0$. In this paper we will study the properties of dynamical systems in kappa deformed space-time and analyse their singularities and comment about the phase transitions they undergo.

In section \ref{sec2} we will briefly go through the theory of $\kappa$-deformations stating the main results without proofs. In \ref{sec3} we discuss the main results from singularity theory defining all terms that will be used later as well as the theorems that will be used. We will follow Arnold's notation throughout the discussion. A more detailed account can be found in Arnold's work \cite{Arnold1988,Arnold1992,Arnold1998,Arnold2012}. In section \ref{sec4} we will apply the results of singularity theory on the Hamiltonian of a particle in quartic potential well in a $\kappa$-Minkowski space showing phase transitions unique to these systems. In sections \ref{sec5},\ref{sec6} we will review methods from the theory of oscillatory integrals, which again comes under the study of singularities, this is being reviewed with the motivation to study wave packets in $\kappa$-minkowski space. We do this in section \ref{sec7}
where we apply the results developed in sections \ref{sec5},\ref{sec6} to find the decay behavior of wave packets. We find that on the cusp catastrophe the wave packets undergo anomalous decay. In section \ref{sec8} we reestablish our calculations of wave packet decay via an analytical approach using Pearcey integrals. Finally in section \ref{sec9} we give some concluding remarks.

\section{$\kappa$-deformed spacetime}\label{sec2}
We will introduce the theory of $\kappa$-deformation on Euclidean space \cite{Meljanac2006}, the corresponding theory for Minkowski space is similar; the difference being that while raising and lowering indices we now have to use the Minkowski metric. The commutator algebra for a $\kappa-$ deformed Euclidean space is 
\begin{equation}\label{eq1}
     [\hat{x}_i,\hat{x}_j] = 0
\end{equation}
\begin{equation}\label{eq2}
    [\hat{x}_0,\hat{x}_i]=ia\hat{x}_i .
\end{equation}
The above two equations can be written in covariant fashion as 
\begin{eqnarray}\label{eq3}
    [\hat{x}_\mu,\hat{x}_\nu]=iC_{\mu \nu}^{\lambda}\hat x_\lambda=i(a_\mu\delta^\lambda_\nu-a_\nu \delta^\lambda_\mu)\hat{x}_\lambda
\end{eqnarray}
where $a_0:=a:=1/\kappa$ and $a_i=0$ we take the standard convention that Latin indices run from one to $n$ and Greek indices run from zero to $n$. We will now realize the non-commutative spacetime coordinates with commutative coordinates $x_\mu$ and derivatives $\partial_\mu$. The most general ansatz with which this is done is the following. 
\begin{align}
    \hat{x}_i &=x_i\varphi(A) \label{eq4}\\
    \hat{x}_0 &=x_0\psi(A)+iax_i\partial_i\gamma(A)\label{eq5},
\end{align}
where $A=-ia\partial_0$ and $\varphi,\psi,\gamma$ are as of now arbitrary functions of $A$. To get the relation between them we put them back in  Eq.~\ref{eq1} and Eq.~\ref{eq2} and we get
\begin{align}
    \frac{\varphi'}{\varphi}\psi &=\gamma-1 \label{eq6}
\end{align}
Now, noncommutative coordinates strictly act as operators, since they involve functions of time derivatives. To map them to commutative coordinates we first define the constant function annihilated by $\partial_0$ as the vacuum $\ket{0}:=1$ and then define the image of a map of noncommutative coordinates in commutative space as 
\begin{align}
    F(\hat x_{\varphi})\ket{0} :&=F_{\varphi}(x) \label{eq7}
\end{align}
where $\varphi$ denotes the particular realization with which we are working. The full $\kappa$-deformed space-time not only consists of the noncommutative coordinates and their algebra but also has the generators of rotations, thus making it into a $\kappa$-deformed Lorentz algebra. We then extend this algebra to include the translations as well and we finally arrive at a $\kappa$-deformed Poincare algebra. We proceed in the same way as before by assuming a general structure for the generators of rotations with commutative coordinates and derivatives and get relations between the arbitrary functions. It is interesting to note that the $\kappa$-deformed Poincare algebra is a deformed Hopf algebra with an undeformed algebra sector and a deformed coalgebra sector. The following are the forms which the generator of rotation and the Dirac derivatives take under a particular realization for $\varphi$
\begin{align}
    M_{ij} &= x_i \partial_j - x_j \partial_i \label{eq8}\\
    M_{i0} &= x_i \partial_0 \varphi \frac{e^{2A} - 1}{2A} - x_0 \partial_i \frac{1}{\varphi} + i a x_i \nabla \frac{1}{2\varphi} - i a x_k \partial_k \partial_i \frac{\gamma}{\varphi} \label{eq9}\\
    D_i &= \partial_i \frac{e^{-A}}{\varphi} \label{eq10}\\
    D_0 &= \partial_0 \frac{\sinh A}{A} - i a \nabla \frac{e^{-A}}{2\varphi^2} \label{eq11}
\end{align}
As we stated earlier, the deformation is in the coalgebra sector and therefore we have a twisted coproduct given by
\begin{align}
    \Delta_\varphi (M_{ij}) &= M_{ij} \otimes I + I \otimes M_{ij} = \Delta_0 (M_{ij}) \label{eq12}\\
    \Delta_\varphi (M_{i0}) &= M_{i0} \otimes I + e^A \otimes M_{i0} + i a \partial_j \frac{1}{\varphi(A)} \otimes M_{ij} \label{eq13}
\end{align}
Another important aspect is the dispersion relation that connects energy and momentum, and it is given by 
\begin{align}
    \frac{4}{a^2} \sinh^2 \left( \frac{aE}{2c} \right) - p_i^2 \frac{e^{-aE/c}}{\varphi^2(aE/c)}\notag \\+ \frac{a^2}{4} \left[ \frac{4}{a^2} \sinh^2 \left( \frac{aE}{2c} \right) - p_i^2 \frac{e^{-aE/c}}{\varphi^2(aE/c)} \right]^2 &= m^2 c^2.\label{eq14}
\end{align}
We will now onward work with the particular realization $\varphi(A)=e^{-A}$ and $\psi (A)=1$ that gives us 
\begin{align}
    \hat{x}_i&=x_ie^{-A} \label{eq15}\\
    \hat{x}_0 &=x_0 \label{eq16}
\end{align}
Using Eq.~\ref{eq14},Eq.~\ref{eq15},Eq.~\ref{eq16} we can derive the Hamiltonian of a particle moving in one dimension in a general potential valid upto first order in the deformation parameter `$a$' as \cite{Verma2018}
\begin{align}
    H(p,x) &= \left( \frac{1 + acm}{2m} \right) p^2 + V \left( x + \frac{ap}{2mc} \right). \label{eq17}
\end{align}
We will now exploit the fact that in $\kappa-$ deformed spacetime the potential gets a momentum dependent term. Before examining that aspect, we review some concepts from singularity and catastrophe theory.

\section{Singularity theory}\label{sec3}
\subsection{function germs and classification of singularities}
We will largely be following the terminologies and methods of V. Arnold's work on singularity theory\cite{Arnold1988,Arnold1992,Arnold1998}. We will now look at some of the definitions and terms that will be used later
\begin{definition}
\textbf{Function germ:} Given a topological space $(X,\tau_{X})$ and two functions $f,g$ defined on $X$ we define an equivalence relation between them at a given point $x\in X$ as $f\sim g \iff \exists  \ \ U\subset X \ \ \ \text{such that } \ \ \ f|_{U}=g|_{U}$ where $U$ is a neighborhood around $x$. We then define the function germ of $f$ $\mathcal{O}_x(f)$ as the equivalence class $[f]$ generated by $f$ i.e.,
$$\mathcal{O}_x(f)=\{g\in F(X): g\sim f\}$$
 where $F(X)$ is the set of all functions defined on $X$
 \end{definition}
 Now there can be multiple distinct function germs at a point. In singularity theory, we are interested in the collection of equivalent function germs at the critical point of a function. We will now define this notion and then define what exactly we mean by a singularity. Without loss of generality, we consider our critical point to be 
$0\in \mathbb{C}$.
\begin{definition}
    \textbf{Equivalence of critical points:}
Consider $\mathcal{O}_n$, the set of all holomorphic function germs defined at $0\in \mathbb{C}^n$. Two function germs at zero are said to be equivalent if one is
taken into the other by a biholomorphic change of coordinates that keeps the
point zero fixed. Two critical points are said to be equivalent if the function germs, after being translated to the origin, are equivalent.
\end{definition}
\begin{definition}
    \textbf{Singularity:} The notion of equivalence of function germs defines an equivalence class at a critical point. This equivalence class of the function germ defined at the critical point is called a singularity.
\end{definition}
\subsubsection{Classification of singularities}
The most influential work of V. Arnold is that of classifying singularities of low modality\cite{Arnold1998}. An interesting point to note is that the classification of simple singularities is very similar to the classification of semi-simple Lie groups, reflection groups, braid groups, etc. Arnold's classification of simple singularities is given below with the type of singularity and their normal forms given. We have two infinite series and three exceptional families.
\begin{table}[htpb]
    \centering
    \renewcommand{\arraystretch}{1.5} 
    \begin{tabular}{ccccc}
        \toprule
        $A_k, k \ge 1$ & $D_k, k \ge 4$ & $E_6$ & $E_7$ & $E_8$ \\
        \midrule
        $x^{k+1}$ & $x^2y + y^{k-1}$ & $x^3 + y^4$ & $x^3 + xy^3$ & $x^3 + y^5$ \\
        \bottomrule
    \end{tabular}
    \caption{Simple singularities of modality $\le 2$}
    \label{tab:simple_singularities1}
\end{table}
\subsection{Versal deformation of singularities}
\begin{definition}
    \textbf{Deformation}: A deformation with the base parameter space $\Lambda=\mathbb{C}^l$ of the germ $f:(\mathbb{C}^n,0)\to \mathbb{C}$ is the germ at zero of a smooth map $F:(\mathbb{C}^n\times\mathbb{C}^l,0)\to \mathbb{C}$ such that $F(x,0)\equiv f(x)$
\end{definition}
Two deformations $F'$ and $F$ of a germ $f$ are called equivalent if $$F'(x,\lambda)=F(g(x,\lambda),\lambda)$$ where $\lambda\in \mathbb{C}^l$ and $g:(\mathbb{C}^n\times \mathbb{C}^l,0)\to \mathbb{C}$ is a smooth germ with $g(x,0)=x$ (a family of diffeomorphisms depending on the parameter $\lambda\in \mathbb{C}^l$). A deformation $F'$ is said to be induced from $F$ if $$F'(x,\lambda)=F(x,\theta(\lambda))$$ where $\theta:(\mathbb{C}^l,0)\to (\mathbb{C}^l,0)$ is a smooth mapping of the bases.
\begin{definition}
    \textbf{Versal deformation: } A deformation $F$ is called versal if every other deformation is equivalent to a deformation induced from $F$.
\end{definition}
\begin{definition}
    \textbf{Miniversal deformation: } A versal deformation is called miniversal if the dimension of the base parameter space is the smallest possible
\end{definition}
We will work exclusively with miniversal deformation, but in order to do so, we need a formula or method to find the miniversal deformation of function germs. This is given by the following theorem.
\begin{theorem}
    A miniversal deformation of a function germ $f(x)$ is given by 
    \begin{equation}
        F(x,\Lambda)=f(x)+\sum_{i=1}^{\mu}\lambda_i\varphi_i(x)
    \end{equation}
    where $\varphi_i(x)$ forms a basis for the local algebra $Q_f$ of the function germ $f$ and $\mu$ is the dimension of the local algebra considered as a $\mathbb{C}$ module.
\end{theorem}
\begin{example}
Consider the function germ $f(x)=x^3$ that has a critical point at zero. We will first find its local algebra. 
$$Q_f=\mathbb{C}[x]/I_{\nabla f}=\mathbb{C}[x]/\langle x^2 \rangle=\{a+bx  \ \ | \ \ a,b\in \mathbb{C}\}$$ 
It is clear that a simple basis for its local algebra is $\{1,x\}$
and therefore its miniversal deformation is $F(x,\Lambda)=x^3+\lambda_1x+\lambda_2$.
\end{example}
\subsection{Bifurcation sets}
The bifurcation set of a singularity contains very important information about the singularity, and by studying this set, we can understand a lot about the nature of the singularity.
\begin{definition}
    The level bifurcation set (or the discriminant) of the singularity $f$ is the germ of the surface $E$ formed in the base of its miniversal deformation by the singular values of the parameter $A$.
\end{definition}
This means that the bifurcation set of a singularity is a surface formed by the values of the parameters for which zero is a critical value for $F(x,\lambda)$. We will explain this with an example. 
\begin{example}
    Consider the function germ $f(x)=x^3$ its miniversal deformation was found in example one as $F(x,\Lambda)=x^3+\lambda_1 x+\lambda_0$ its bifurcation set is those values of $\lambda_1,\lambda_0$ for which the polynomial $x^3+\lambda_1 x+\lambda_0$ has multiple roots. That means we solve $F(x,\Lambda)=0$ and $F'(x,\Lambda)=0$ solving them and eliminating $x$ we get 
    $$27\lambda_0^2+4\lambda_1^3=0$$
\end{example}
\begin{figure}[h]
    \centering
    \includegraphics[width=0.9\linewidth]{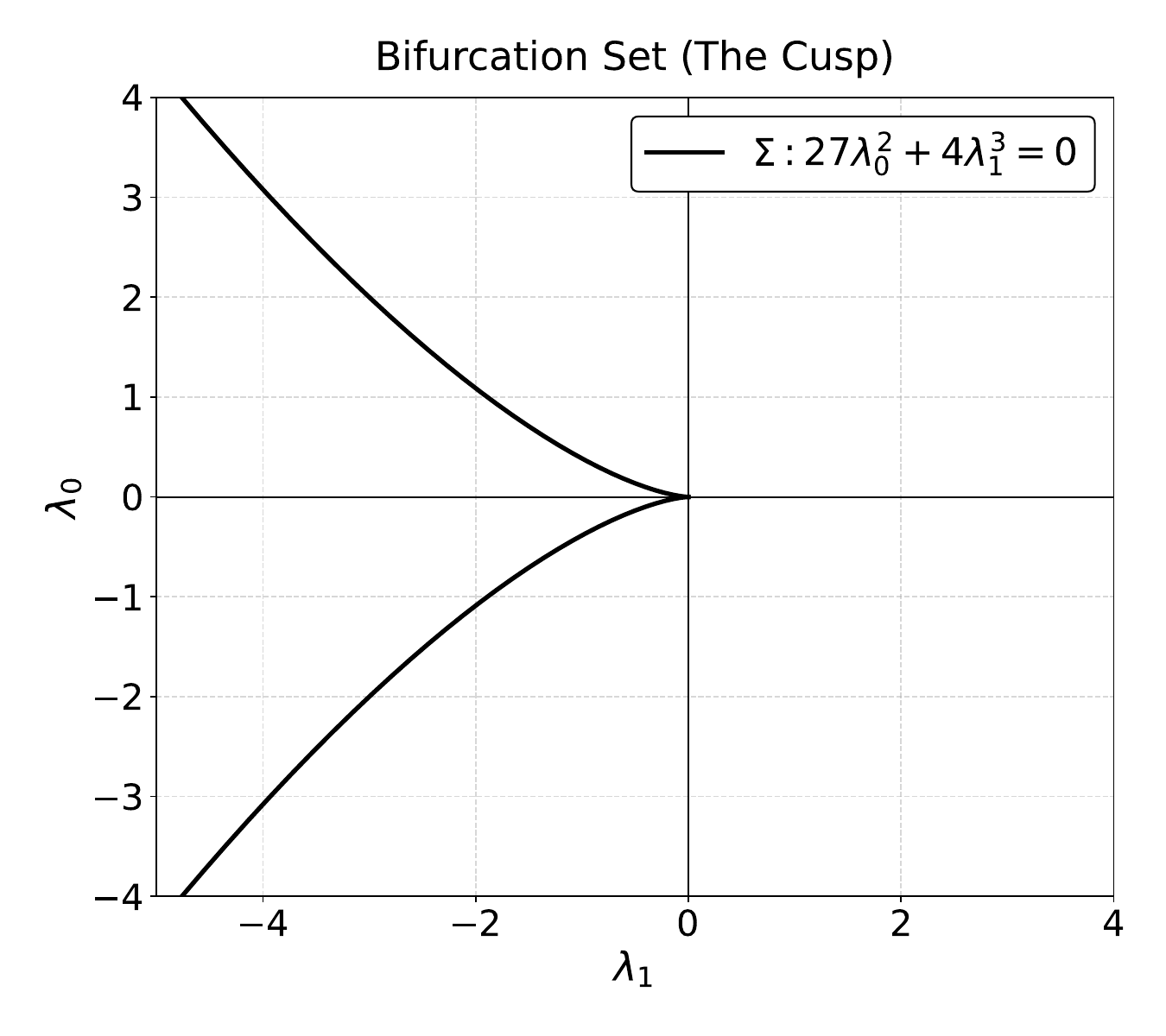}
    \caption{Bifurcation set for $A_2$ singularity}
    \label{fig:placeholder2}
\end{figure}
The bifurcation set also divides the parameter space into two regions. Within a given region, the nature of the roots must remain unchanged. Therefore, by analyzing the nature of the roots at a convenient point in each of the two regions, we can determine how many solutions exist in each region. Towards the left, we have three distinct roots, but towards the right we have only one real root, and at the cusp we have three same roots. On the bifurcation set we have in general multiple roots, one distinct and the other two identical. There is another type of bifurcation set that is useful, called a function bifurcation set. To define the function bifurcation set we need the concept of a Morse function and the concept of a truncated miniversal deformation.
\begin{definition}
    \textbf{Truncated miniversal deformation:} It is the miniversal deformation of a function germ such that the deformed function lies within the class of functions that vanish at origin.
\end{definition}
\begin{definition}
    \textbf{Morse function:} A function $h(x)$ is said to be a Morse function on the set $U$ if it has only
nondegenerate critical points on $U$ and its critical values at these points are all distinct.
\end{definition}
\begin{definition}
    \textbf{Function bifurcation set:}The function bifurcation set for $f$ is the germ at zero of the hypersurface $\Sigma$ formed in  the base $\Lambda'\cong \mathbb{C}^{\mu-1}$ of a truncated miniversal deformation
by the values of the parameter $\lambda'$ such that there does not exist a neighborhood around origin for which the function $x \mapsto F'(x,\Lambda')$ is a
Morse function.
\end{definition}
 Since the hypersurface $\Sigma$ must be such that the parameters of deformation do not make the deformation a Morse function, it must either have degenerate (which means at the critical point at least the first and second derivatives must vanish) or the critical values must coincide. A better appreciation for this comes from an example.
\begin{figure}[h]
    \centering
    \includegraphics[width=0.9\linewidth]{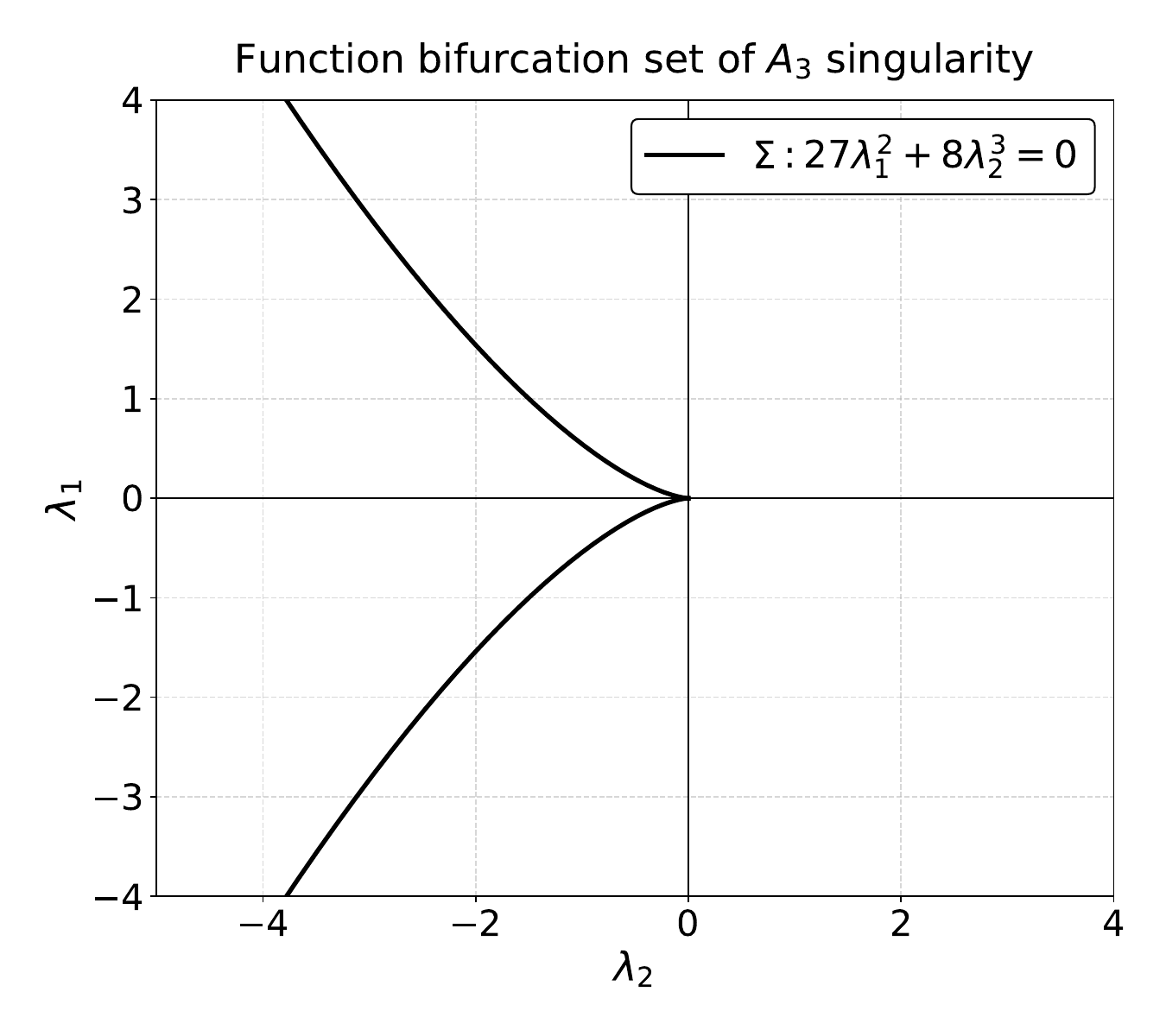}
    \caption{function bifurcation set of $A_3$ singularity}
    \label{fig:placeholder1}
\end{figure}

\begin{example}
    Consider the function germ $f(x)=x^4$. The truncated miniversal deformation of this function germ is given by $F(x,\Lambda)=x^4+\lambda_2x^2+\lambda_1x$. For this to have degenerate critical points, its first derivative and second derivative must be zero, solving the two equations by eliminating $x_c$ gives $27 \lambda_1^2 + 8\lambda_2^3 = 0$. This is called caustic. Now, critical values coincide when it is an even function, i.e., when $\lambda_1=0$ this defines the magnus stratum as shown in the figure.
\end{example}
In the following section, we give a very brief account of the theory of Lagrangian singularities.
\subsection{Lagrangian singularities.}
Consider the phase space as $\mathcal{M}=\mathbb{R}^{2n}$ with coordinates $(p_1,...,p_n;q_1,...,q_n)$ equipped with the standard symplectic two-form  $\omega=\sum_i dp_i \wedge dq_i$. We will define the concepts of Lagrangian submanifolds and Lagrangian projections onto this standard phase space.
\begin{definition}
    \textbf{Lagrangian Submanifold:} A Lagrangian submanifold $\mathcal{L}\subset \mathcal{M}$ is a manifold of half the dimension (i.e., $n$) of the phase space such that the two-form restricted to this submanifold induces the null form that is $\omega|_{\mathcal{L}}=0$
\end{definition}
\begin{definition}
\textbf{Lagrangian fiber bundles}
    A fiber bundle $\pi : E^{2n} \to B^n $ is said to be Lagrangian if the space $E^{2n}$ is equipped
with a symplectic structure and the fibers are Lagrangian submanifolds. The map $\pi$ is called a Lagrangian projection.
\end{definition}
The standard Lagrangian fiber bundle is defined as $E^{2n}=\mathbb{R}^{2n}$ and $B^n=\mathbb{R}^n$
\begin{definition}
    \textbf{Lagrangian map:} the Lagrangian projection $\pi$ restricted to a Lagrangian submanifold is called a Lagrangian map $$\pi |_{\mathcal{L}}: \mathcal{L}\to \mathbb{R}^n$$
\end{definition}
The Lagrangian map is a smooth map between two manifolds of equal dimension. We can now define the critical points of this map as the points $x\in \mathcal{L}$ such that the Jacobian determinant of the differential $d(\pi |_{\mathcal{L}})_x$ is zero
\begin{definition}
    \textbf{Caustic:} The image set of the critical points of the Lagrangian map under the projection is called caustic.
\end{definition}
\section{Singularity theory applied to $\kappa$-deformed quartic potential}\label{sec4}
Now we are ready to explore the implications of the singularity theory applied to a $\kappa-$deformed quartic potential. The standard undeformed quartic potential is given by 
\begin{equation}
    H(p,x)=\frac{p^2}{2m}+x^4
\end{equation}
We now apply a truncated miniversal deformation to the system to  get the deformed Hamiltonian to be 
\begin{eqnarray}
    H(p,x;b,c)=\frac{p^2}{2m}+x^4+bx^2+cx
\end{eqnarray}
Now in addition to a truncated miniversal deformation, we add the $\kappa-$ deformation to the system, hence by using Eq.~\ref{eq17} we get
\begin{equation}
    H_a(p,x,b,c)=\frac{p^2}{2m_{eff}}+(x+\alpha p)^4+b(x+\alpha p)^2+c(x+\alpha p)\label{eq21}
\end{equation}
where $\alpha=a/2mc$. We first analyze whether the equilibrium of the system is changed due to the effect of $\kappa$-deformation. The dynamic state equilibrium is attained when both $\dot{q}=0$ and $\dot{p}=0$. From Hamilton's equations, we get 
$\partial \mathcal{H}/\partial x = 0$ and 
$\partial \mathcal{H}/\partial p = 0$. This, using Eq. 17, leads to the following:
$$ V'\left(x+\frac{ap}{2mc}\right)=0,  
\,\, \mathrm{and} \,\, \frac{p}{m_{eff}}
+V'\left(x+\frac{ap}{2mc}\right)\frac{a}{2mc}=0$$
From the first of these two, we get $p=0$, which in turn, from the second relation, yields 
$V'(x)= 0$.

We therefore see that the point of equilibrium is not changed. In fact, an argument can be made by using Arnold's singularity theory according to which the type of singularity is unchanged under diffeomorphism. The change brought about by a $\kappa-$ deformation in the Hamiltonian is nothing but a linear shift, which is a canonical transformation, which is nothing but a symplectomorphism and therefore a diffeomorphism. So the singularity at the critical point remains invariant hence the local behaviour of the function germ will not change, therefore we get the same points of equilibrium as before.
\subsection{Caustics of the level sets of $\kappa-$deformed quartic potential.}
Now let us look at an arbitrary energy state with energy $E$. the relevant Lagrangian submanifold is
$$\mathcal{L}(E,p,q)=\{(p,q) \ | \ \mathcal{H}_a(p,q)=E\}$$ This submanifold is Lagrangian because by introducing the constraint of constant energy we reduce the dimension of the phase space from two to one, so the dimension of the submanifold is exactly half of that of the full phase space. One can also verify that the two form restricted to this submanifold induces a null form therefore making it a Lagrangian submanifold. The caustic is the map of the critical points of the Lagrangian map. In this case, we project the level set to the configuration space, and the critical points are those points at which the slope becomes vertical, that is, the points where 
$\partial\mathcal{H}/\partial p = 0$, leading to
$$ \frac{p}{m_{eff}}+V'\left(q+\frac{ap}{2mc}\right)\frac{a}{2mc}=0$$ To analyse the catastrophe structure, we study the function bifurcation set of the cusp potential under $\kappa$-deformation which is 
$$V(x+\alpha p)=(x+\alpha p)^4+b(x+\alpha p)^2+c(x+\alpha p)$$ So the Lagrangian submanifold is given by \footnote{Here even though the argument is expanded in first order at $\alpha$ we will keep all higher order terms in the explicit expression for $V(x+\alpha p)$ this is because we are working in the high $\alpha p$ regime and so we will keep higher powers of this but neglect $O(\alpha^2)$ terms in the argument.}
\begin{equation}
    \frac{p^2}{2m_{eff}}+(x+\alpha p)^4+b(x+\alpha p)^2+c(x+\alpha p)=E\label{eq22}
\end{equation}
To make the study simpler, we look at the submanifold of the phase space when $x=0$, that is, we look at the `momentum' space part of the phase space keeping $x=0$ fixed. In this way, we can focus on the topological features of this submanifold. Following in this pattern, Eq.~\ref{eq22} becomes
\begin{equation}
    \frac{p^2}{2m_{eff}}+\alpha^4p^4+b\alpha^2p^2+c\alpha p=E\label{eq23}
\end{equation}
Now, we define
$$\lambda_2=\frac{1}{\alpha^4}\left(\alpha^2 b+\frac{1}{2m_{eff}}\right) \,\, \mathrm{and} \,\, \lambda_1=\frac{c}{\alpha^3},$$ 
Eq.~\ref{eq23} becomes the standard miniversal deformation of the $A_3$ singularity in the variable $p$. Now we can refer back to our Example 3 to get the function bifurcation set as
\begin{equation}
    27\lambda_1^2+8\lambda_2^3=0
\end{equation}
By studying this bifurcation set, we see that if $\lambda_2>0$ there is no possibility of a transition in the parameter space which crosses the bifurcation curve, so we must have $\lambda_2<0$, which gives us an upper bound on $b$. 
\begin{equation}
     b<\frac{-1}{2\alpha^2m_{eff}}\label{eq25}
\end{equation}
This gives a crucial bound on $b$ for which the transition exists. In the limit as $\alpha \to0$ we see that $b \to-\infty$ so we can conclude that this kind of topological phase transition is induced only in a $\kappa$-deformed system.

We see that $\kappa$-deformed systems have interesting transitions that occur in the momentum space. The Lagrangian submanifold we considered was the level set with energy $E$ and we see from Eq. 25 that for a particular choice of the control parameter $b$ (depending on the $\kappa$-deformation constant) we get a cusp catastrophe in momentum space. This means that  in the family of level sets parametrized by $b,c$ there exist multiple momentum branches (as seen from the bifurcation set), which happens only when $b$ crosses the threshold given by Eq. 25. An easier way to visualize this is to look at the quartic function $x^4+bx^2+cx$ that exhibits a double well structure only if $b<0$. If we compare this with our $\kappa$-deformed system it tells us that even though $b<0$ is enough to have a double well structure in the configuration space, to get a double well structure in the momentum space we must have that $b$ is less than a large (since $\alpha$ will be small) negative number given by $\alpha$. Another interesting observation is that once we find such a $b$, i.e. the value of the tuning parameter at which the transition occurs, inequality \ref{eq25} can be inverted to obtain a lower bound for $\alpha$. This will imply that the $\kappa$-deformation parameter is non zero meaning that space time is indeed noncommutative. To see how these transitions affect the system, one has to study the topology of the energy level manifold. If this topology changes, we can call the transition a topological transition.

Topological transitions take place when changing a control parameter changes the topology of the Manifold, we claim that the transition from $\lambda_2>0$ to $\lambda_2<0$ marks a topological transition. To prove this, we calculate the Morse index of the critical point $(0,0)$ of the level manifold $\mathcal{H}_{\alpha}(x,p,\lambda_2,\lambda_1)=E$ and show that it changes as we vary the parameter $\lambda_2$.

The Morse index is defined as the number of negative eigenvalues of the Hessian at the critical point. The Hessian can be calculated from Eq.~\ref{eq21}. We get
\begin{align}
    H=\begin{pmatrix}
\lambda'_2 & 2b\alpha \\
2b\alpha & 2b
\end{pmatrix}
\end{align}
where $\lambda_2'=2\lambda_2\alpha^4$
eigenvalues are given by
\begin{equation}
    \lambda=\frac{\lambda'_2+2b\pm\sqrt{(\lambda'_2+2b)^2-8b\lambda'_2+16b^2\alpha^2}}{2}
\end{equation}
for $\lambda'_2>0$ say $\lambda'_2=1/m_{eff}$ we get the eigenvalues as $\lambda=0,\lambda'_2$ so the Morse index $k=0$. Now, at $\lambda'_2=0$ we have $b=-1/2\alpha^2m_{eff}$. Using these values, we find that one of the eigenvalues is positive while the other is negative, which gives us a Morse index $k=1$. The meaning of having a Morse index of zero is that the critical point is a minima while if it is one, then we can say that the critical point is a saddle while if it is two, then it is a maxima. This change in the Morse index clearly reflects the change in the function as it changes from a single well to a double well and the point $p=0$ which was a minima changes into a saddle and then into a local maxima. Now, to see how the level sets are topologically different, we can use the following theorem.
\begin{theorem}
Let $M$ be a smooth $m$ dimensional manifold and let $f$ be a morse function defined on it. Consider the sublevel sets $M^a:=\{x\in M \ : f(x)\leq a\}$ and $M^b_a=\{x\in M \ : a\leq f(x)\leq b\}$, then under the standard compactness assumptions:
\begin{enumerate}
    \item The sublevel sets $M^a$ is diffeomorphic to $M^b$ if $M^b_a$ does not contain any critical points
    \item If there is one critical point $x$ in the interior of $M^b_a$ with morse index $k$, then $M^b\cong M^a \cup H^m_k$, where $H^m_k:=D^k\times D^{m-k}$ is a $k$-handle attached along its attachment boundary $\partial_0 H^m_k=S^{k-1}\times D^{m-k}$..
\end{enumerate}. 
\end{theorem}
In our case, the smooth manifold $M$ is the phase space and the function $f$ is our Hamiltonian. For the theorem to hold, we need the function $f$ to be a Morse function, and to show that a function is a Morse function we need to prove that its Hessian is non degenerate. For our Hamiltonian we have calculated the Hessian and from that we can see that it is indeed a morse function.

For our system, to show that the topology of the level sets changes, we first look at the case where $\lambda_2>0$, which is a single well potential with only one critical point $p=0$. The energy at this point is $E=0$. Now, for any $a,b$ such that $0<a\leq b$, $M^b_a$ does not have any critical point, therefore, all sublevel sets $M^a$ are diffeomorphic to other sublevel sets. Next, if we change the parameter $\lambda_2$ to $\lambda_2<0$, new energy states become accessible with negative energies, indicating bound states. This structure becomes a double well and if we look at the sublevel sets $M^a$ and $M^b$ with $a<0<b$, then $M^a$ has two sets of connected components, meaning that their zeroth homotopy set $\pi_0$ has cardinality two. On the other hand, if we look at $M^b$ which has $b>0$, this makes it a connected space, and therefore cardinality of its zeroth homotopy set is one. Therefore, we can conclude $|\pi_0(M^a)|\neq |\pi_0(M^b)|$ and which makes them topologically distinct (they are not homeomorphic). 

To actually see how this non-trivial topology manifests in real space-time, we study the wave packets and their asymptotics by using Arnold's theory of oscillatory integrals. This gives us the leading order behaviour of the wave-packets and we will see that the behaviour is drastically different from the wave packets in commutative spaces.

\section{Basics of Oscillatory integrals}\label{sec5}
In this section, we will first review the theory of oscillatory integrals. We will state a few theorems and some definitions, and for details, the readers are suggested to refer to Arnold's book\cite{Arnold1988}.

An oscillatory integral is an integral of the form
\begin{equation}
    I(\tau)=\int_{\mathbb{R}^n} e^{i\tau f(x)}\phi(x) \, dx_1dx_2...dx_n
\end{equation}
One can study the asymptotic behavior of this integral as $\tau$ goes to infinity. According to the principle of stationary phase, the values of $x$ that actually contribute the most in this oscillatory integral are the critical points of the phase $f(x)$. It turns out that the complete order behavior of the integral for large $\tau$ can be understood by studying the critical points of the phase, especially the degenerate singularities. A general asymptotic expansion for this integral is 
\begin{equation}   
I(\tau)\sim\exp \left( i\tau f(x_0) \right) \sum_{\alpha} \sum_{k=0}^{n-1} a_{k,\alpha}(\varphi) \tau^{\alpha} (\ln \tau)^k \, , \, \tau \to +\infty \label{eq29}
\end{equation}
given that the phase is an analytic function in a neighborhood of the critical point $x_0$ and that the amplitude has a support that is concentrated in a sufficiently small neighborhood around the critical point. Here, the index $\alpha$ runs over an arithmetic progression, where it takes negative rational values, and the coefficients are generalized functions that depend only on the amplitude and not on the phase. There are several theorems and techniques that one can use to get the values of these indices and the coefficients, which we will look at next.

One of the most influential works of Arnold is his classification of singularities as we have reviewed in section 3.11. If the phase is of one of the normal forms of these simple singularities, then we can predict the order of the leading term in the asymptotic expansion of the oscillatory integral. 
\begin{theorem}
    Each degenerate singularity in the phase of the integral contributes a term of order $\tau ^{\beta -n/2}$ where $\beta$ is called the oscillation index, which depends on the singularity type, and is summarized in the table below.
\end{theorem}
\begin{table}[htpb]
    \centering
    \renewcommand{\arraystretch}{1.5} 
    \caption{Oscillation indices of simple singularities of modality $\le 2$}
    \begin{tabular}{ccccc}
        \toprule
        $A_k, k \ge 1$ \,\,& $D_k, k \ge 4$ \,\,& $E_6$ \,\,& $E_7$ \,\, & $E_8$ \\
        \midrule
        $\dfrac{k-1}{2k+2}$ & $\dfrac{k-2}{2k-2}$ & $\dfrac{5}{12}$ & $\dfrac{4}{9}$ & $\dfrac{7}{15}$\\
        \bottomrule
    \end{tabular}
    \label{tab:simple_singularities2}
\end{table}
We will next look at a few key definitions needed to state and use the main theorem and then look at the concept of Newton diagrams in terms of which the theorem is formulated. 
\begin{definition}
    \textbf{Index set:}The index set of an analytic phase at a critical point is the set of all numbers $\alpha$ possessing the property: for any neighbourhood of the critical point there is an amplitude with support in the neighbourhood for which in the asymptotic series Eq.~\ref{eq29} there is a number $k$ with the property that the coefficient $a_{k,\alpha}$ is not equal to zero.
\end{definition}
\begin{definition}
    \textbf{Oscillation index of a critical point:} It is the maximal element of the index set denoted by $\beta$
\end{definition}
\begin{definition}
    \textbf{Multiplicity of the oscillation index:} It is the maximal number $k$ possessing the property: for any neighbourhood of the critical point there is an amplitude with support in the neighbourhood for which in the asymptotic series Eq.~\ref{eq29} there is a number $k$ with the property that the coefficient $a_{k,\alpha}$ is not equal to zero.
\end{definition}
In the next section, we will introduce the concept of Newton diagrams and Newton polyhedrons and define what we mean by \textit{remoteness} of the Newton diagram in terms of which the main theorem is formulated.

\section{Newton Diagrams and Newton Polyhedrons}\label{sec6}
Let $\mathbb{Z}^n_+$ denote all the integer lattice points in the positive orthant of the space $\mathbb{R}^n$ and let $S$ be an arbitrary subset of $\mathbb{Z}^n_+$ we define the Newton polyhedron of this space $\Gamma(S)$ below
\begin{definition}
    \textbf{Newton polyhedron:} The Newton polyhedron $\Gamma(S)$ of $S\subset \mathbb{Z}^n_+$ is the convex hull of the union of subsets of $\mathbb{R}^n$ consisting of points that have all their coordinates greater than the corresponding coordinate of a particular element 
    $s\in S$, i.e.,
    \begin{align}
        \Gamma(S) &=\text{Con}(\bigcup_{s\in S} R_s) \\
        R_s &=\{x\in \mathbb{R}^n \  |x_i\ge s_i \ \ 0\le i\le n\}
    \end{align}
    where $x_i$ and $s_i$ denote the $i^{\text{th}}$ coordinate of $x$ and $s$
\end{definition}
\begin{definition}
    \textbf{Newton diagram:} The Newton diagram of an arbitrary set $S$ is the union of compact faces of its Newton polyhedron denoted by $\Delta(S)$
\end{definition}
\begin{definition}
    \textbf{Remoteness:} The point of intersection of the Newton diagram with the diagonal $x_1=x_2=...=x_n$ is called the center of the boundary of the Newton diagram, remoteness is defined as the additive inverse of its reciprocal. It is denoted by $\beta$. A Newton polyhedron is called remote if the center of the boundary is more than one.
\end{definition}
In our analysis, we are concerned with the Newton diagram of a function $f(x)$ about its critical point. In order to achieve this, we first expand the function as a Taylor series about this critical point, and then we define the support of the Taylor expansion below (which has a different meaning from the support of a function which was assumed on the amplitude of the phase integral)
\begin{definition}
    \textbf{Support of the Taylor expansion of a function:} Let the critical point of the function be $x=0$ and its Taylor expansion be $f(x)=\sum_k a_kx^k$, where $k$ is a multi-index. Then, the support of the function is the collection of $k$ with $a_k\ne 0$  
\end{definition}
Then the Newton diagram of the function $f$ is just the Newton diagram of its support with respect to the Taylor expansion around the critical point. Next we define the concept of the principal part of a function and its non-degeneracy.
\begin{definition}
    \textbf{Principal part of a function} The principal part of a function is the polynomial part of the Taylor expansion, whose monomials have indices on the Newton diagram. The part of the expansion whose monomials have indices on a compact face is denoted by 
    $f_\gamma $ and therefore the principal part is $f_\Delta=\sum_i f_{\gamma_i}$(since the Newton diagram is the sum of compact faces of the Newton polyhedron).
\end{definition}    
\begin{definition}
    \textbf{Nondegeneracy of the Principle part of a function:}
    The principal part of the power series of $f$ with real coefficients (respectively power series with complex coefficients) is $\mathbb{R}$-nondegenerate (resp. $\mathbb{C}$-nondegenerate) if for every compact face $\gamma$ of the Newton polyhedron of the series, the polynomials 
    $\left\{\dfrac{\partial f_\gamma}{\partial x_i}\right\}$ do not have a common zero in $\mathbb{R}^n$ (respectively in $\mathbb{C}^n$)
\end{definition}
We now formulate the main theorem which can be found in the book by Arnold \cite{Arnold1988} as theorem 6.4.

\begin{theorem}
    Let the phase be an analytic function in a neighbourhood of its critical point. Let us suppose that the principal part of the Taylor series of the phase at this critical point is $\mathbb{R}$-nondegenerate and that the Newton polyhedron is remote. Then the oscillation index of the critical point of the phase is equal to the remoteness of the Newton polyhedron.
\end{theorem}
This is a very powerful result, saying that just by drawing out the Newton diagram of the phase we can understand the asymptotic behavior of the oscillatory index. In our case this translates to the fact that by drawing the Newton diagram of the Hamiltonian we can find the decay order of wave packets. Now we are ready to look at the dynamics of the wave packet. We will compare the results we get with that of the standard wave packets in commutative space times.

\section{Wave packet asymptotics}\label{sec7}
\begin{figure}
    \centering
    \includegraphics[width=1\linewidth]{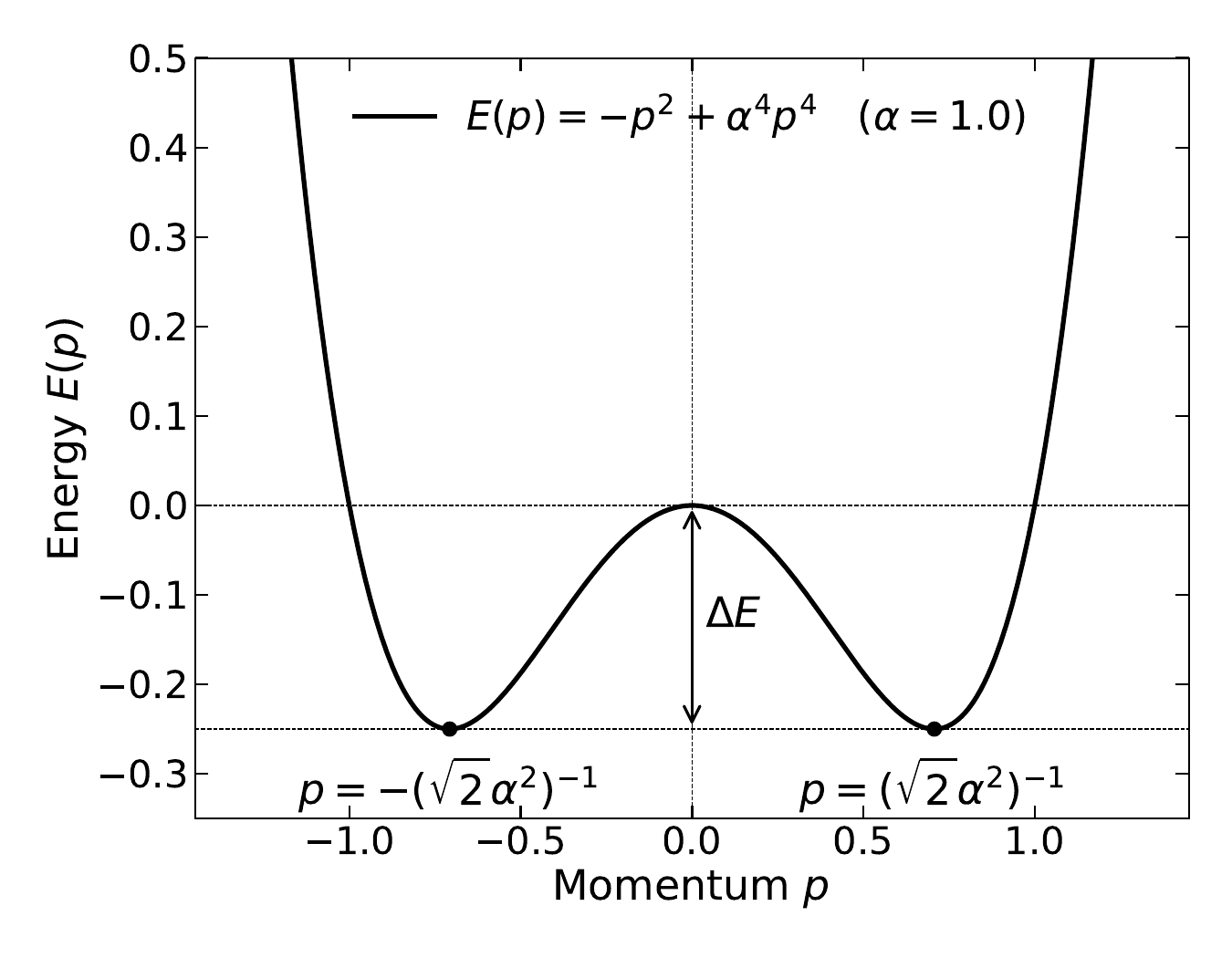}
    \caption{Energy momentum relation with a double well topology. The depth of the double well $\Delta E=1/4\alpha^4$ depends on the noncommutative parameter $\alpha$}
    \label{fig3}
\end{figure}

The standard wave packets in quantum mechanics are just superposition of plane waves with weights given by a function $\phi(k)$ which is a Fourier transform of an initially `well behaved' wave, say, a Gaussian wave and then takes its time evolution by using a dispersion relation $w(k)$. Thus, a wave packet looks like 
\begin{equation}
    \Psi(x,t) = \frac{1}{\sqrt{2\pi}} \int_{-\infty}^{\infty} \phi(k) \exp\Big( i \big( kx - \omega(k)t \big) \Big) dk
\end{equation}
In our discussions, we will study the asymptotics of this wave packet by using the standard dispersion relation of a free particle and that of a particle in a momentum double well set up. Since in both these cases the Hamiltonian will only be a function of $p$, it will commute with the momentum operator, and therefore in the time evolution part we can replace the Hamiltonian operator with the function $H(p)$. 
\begin{equation}
   \Psi(x,t) = \frac{1}{\sqrt{2\pi\hbar}} \int_{-\infty}^{\infty} \phi(p) \exp\left[ \frac{i}{\hbar} \big( px - H(p)t \big) \right] dp
\end{equation}
Recall that our $\kappa$-deformed Hamiltonian is given by Eq.~\ref{eq21}. We take the case of a symmetric double well so that the linear term can be omitted. We now take an approximation so that we can justify studying the system at $x=0$. The assumption is that we will be studying the high momentum limit of the system where $\alpha p>>x$ or a universe where $\alpha$ is sufficiently large. In this manner, one can come up with the energy momentum relation which is valid in the high momentum regime. That is $E(p)\approx H(p,0)$ we will use this dispersion relation in the upcoming discussions. Thus, we get
\begin{equation}
    E(p)=\frac{p^2}{2m_{eff}}+\alpha^4p^4+b\alpha^2p^2
\end{equation}
Now, to obtain a double well structure in momentum space, we use Eq.~25 to define $b\alpha^2+1/2m_{efff}=-1$, to get the final dispersion relation as (see Fig.~\ref{fig3})
\begin{equation}
    E(p)=-p^2+\alpha^4p^4
\end{equation}
In fact, such dispersion relations have already been studied in the context of Higher derivative field theories\cite{Lee-Wick}, Lorentz violating effective field theories\cite{eff_field-theory} and quantum gravity models such as Hořava-Lifshitz gravity\cite{Horawa_gravity}. The interesting thing about this dispersion relation is that at low momenta it gives an imaginary energy value, which is referred to as the tachyonic instability. To compensate for this, a fourth order term in p is added so that tachyonic instability can be avoided at higher momenta. \cite{tachyoninstability}.

We will also look at the case where $b\alpha^2+1/2m_{efff}=0$, which exactly corresponds to a point in the caustic (since $c=0$, after solving the function bifurcation set, we find that the point in the caustic corresponding to $c=0$ is $b\alpha^2+1/2m_{efff}=0$). Therefore, we get the deformed dispersion relation as 
\begin{equation}
    E(p)=\alpha^4p^4
\end{equation}
There is one more important context where dispersion relations of the form given in Eqs.35,36 are studied, namely quartic solitons \cite{quarticsolitons,Solitonsreview,Solitonsinmultimode}. Solitons are wavepackets that can propagate without changing their shape and quartic solitons are solitons appearing in optical models which have a dispersion relation dominated by a fourth order term, much like what we have in Eq.~35. Moreover, now physicists have been able to realize pure quartic solitons too (solitons with a purely quartic dispersion relation); therefore, the discussion that follows about the decay behavior of wavepackets can also be applied to such soliton systems, essentially showing that quartic solitons and pure quartic solitons have different decay behavior.

Before starting to apply the methods of oscillatory integrals, there are two technical details that need elaboration. 
\begin{enumerate}
\item  The first is related to the amplitudes. In general, for the study of oscillatory integrals it is required that the phase has a compact support and that it is concentrated in a neighborhood around the critical point. This can be achieved by using a very narrow initial Gaussian (i.e., very small standard deviation) for the weights $\phi(p)$ and centered around the critical point of the phase (in our case, we will study the critical point $p=0$ in both cases). Then we can restrict the region of integration to finite limits (a compact region) in the Fourier transform as momentum from outside this compact set contributes very little 
(this can also be justified using the fact that we are working in the high momentum limit and as such contribution from lower values of momenta will not contribute to the integral). We will assume such technical `tunings' of the amplitude are always possible at least via suitable approximations. In any case the amplitude only contributes to the numerical factor in front of the leading order term in the asymptotic expansion, so we will focus more on the phase of the oscillatory integral.

\item The second technical detail comes because of the use of the Fourier transform in a noncommutative space. If the space is noncommutative it introduces a minimum length scale for the universe and this will effect the uncertainty relations. Also, a proper position space representation in a Hilbert space becomes impossible. A recent paper \cite{fourier_in_NCspace} shows how to deal with this issue by first writing the wave functions in their momentum representation and then by taking a Fourier transform with a modified measure. The plane wave part of $e^{ipx}$ changes and the final Fourier transformation to what is called a `maximal localized space' is given by 
\begin{eqnarray}
\psi(\xi) &=& \sqrt{\frac{2\sqrt{\beta}}{\pi}} 
\int_{-\infty}^{+\infty} \frac{\phi(p)}{(1+\beta p^2)^{3/2}} \times \nonumber \\ 
&& \hspace{48pt} \exp\left(\frac{i\xi \arctan(\sqrt{\beta}p)}{\hbar\sqrt{\beta}}\right) dp
\end{eqnarray}
Here $\beta$ is the constant that appears in the generalized uncertainity principle $\Delta x\Delta y=\dfrac{\hbar}{2}(1+\beta(\Delta p)+\gamma)$ due to the noncommutative nature of spacetime.
This integral is guaranteed to converge, provided that the function $\phi(p)$ belongs to the Schwartz space.
Now we will have to use this expression for our wave packets. We again assume that the technical requirements of the amplitude functions are met. To study the large time asymptotics we will break the $\exp(ipx)$ term or the above modified plane wave expression using Euler's formula (this is done because these plane wave expressions do not have a degenerate critical point and also are not multiplied with large order parameter $t$ and so this will not contribute to the leading order term in the expansion) and write the final wave packet as
\begin{eqnarray}
    \Psi(x,t) = \int e^{-itH(p)/\hbar} \left[A_1(p,x) +i A_2(p,x) \right] dp
\end{eqnarray}
where $A_1,A_2$ are the modified amplitudes that meet all the technical requirements to use the theorem.
\end{enumerate}

We can now fully focus on the phase of the oscillatory integral (which is the same for both terms), which is the Hamiltonian, and we will analyze the large $t$ asymptotics for the critical point $p=0$ in both the commutative and noncommutative setting.

Since the phase is a polynomial function in just one variable, the principal part of this will always be non degenerate and Newton polyhedrons will just be subsets of $\mathbb{R}$ and Newton diagrams will be points in $\mathbb{R}$. For the commutative case, the phase function is $p^2/2m_{eff}$, so the Newton polyhedron will be the subset $[2,\infty)$. The Newton diagram will be the compact face of this (here face means a $n-1$ dimensional subset of the polyhedron) so $\Delta=\{2\}$ and therefore its center of boundary is also $2$ and hence is remote , finally the remoteness of this Newton diagram will be
$\beta=-1/2$ and therefore we get that for large $t$ in the commutative case the wave packet behaves as:
\begin{equation}
    \Psi(x,t) \sim t^{-1/2}
\end{equation}
Now for the noncommutative case our phase is $H(p)=-p^2+\alpha^4p^4$ the support of this is $\{2,4\}$ and hence $\Gamma(H)=[2,\infty)$ and Newton diagram is $\Delta(H)=\{2\}$ hence its remoteness is $\beta =-1/2$ therefore the wave packet decays as
\begin{equation}
    \Psi(x,t) \sim t^{-1/2}
\end{equation}
Now if we look at the case when dispersion relation is $E(p)=\alpha^4 p^4$, the support is the single point $\{4\}$ and therefore its Newton diagram is $\Delta(H)=\{4\}$ and therefore its remoteness is $\beta=-1/4$ and so the wave packet shows the anomalous decay behavior.
\begin{equation}
    \Psi(x,t) \sim t^{-1/4}
\end{equation}
Therefore what it means is that as one tunes the control parameter $b$ it passes through a critical point that we mentioned in Eq.~25. This crossing causes a topological transition in the momentum space as we had proved before. This topological transition is manifested in the decay behavior of wave packets because exactly at the critical point the wave packet shows an anomalous decay behavior which can be used as a test to prove the existence of this topological phase transition. This will also prove the existence of a noncommutative space time, because this kind of behavior is only seen in the $\kappa-$deformed version of the system. In the next section we will find the oscillatory behavior for our system in more direct way by using Pearcey integrals thus verifying techniques used from singularity theory.

\section{Asymptotics of wave packets in $\kappa$-Minkowski space via Pearcey integral asymptotics}\label{sec8}
 Various studies on the Pearcey integral have been conducted from a physical perspective \cite{CACCIATORI2024116477} as well as from a pure mathematics perspective \cite{analytical_formulasofpearceyintegral}. In 
 addition, several investigations have been reported 
  on the asymptotics of Pearcey integrals and their connection to singularity theory  \cite{pearcey_complex}. We shall now obtain the asymptotic behavior of wavepackets in $\kappa$-Minkowski space by using analytical formulas for Pearcey integrals.

For simplicity, we will look at Eq.33 only, leaving the more difficult part of doing it with the modified Fourier transform via Eq.43 for the future. This is done here because we are only interested in the large order behavior of wave packets as $t\rightarrow \infty$ and we expect this to be the same for both cases as asymptotic behavior largely depends on the phase function and not the amplitude. Therefore, starting with a Gaussian envelope for $\phi(p)$ we get an oscillatory integral 
\begin{equation}
    I=A\int^{+\infty}_{-\infty}\exp\left(-\frac{(p-p_0)^2}{4\sigma_p^{2}}+\frac{i}{\hbar}\left(p x-H(p)t\right)\right) \,dp
\end{equation}
where $A$ is a normalization constant. As we can see, this integral is not in the standard form of the Pearcey integral, we will use a change of variables to convert it into the desired form. The phase function is 
\begin{equation}
    \phi(p)=i\frac{(p-p_0)^2}{4\sigma_p}+\frac{px-H(p)t}{\hbar}
\end{equation}
where we take $H(p)=bp^2+\alpha^4 p^4 /4$ and use the change of variable $p=y\sqrt{2i}/\left(\alpha^4t\right)^{1/4}$ and for this case we take that the initial Gaussian envelope is around $p_0=0$, and then the phase function becomes
\begin{equation}
    \phi(y)=\left(\frac{-2}{4\sigma_p\alpha^2\sqrt{t}}-\frac{2bi\sqrt{t}}{\alpha^2\hbar}\right)y^2+\frac{(1+i)x}{\hbar\alpha\, t^{1/4}}y+\frac{y^4}{\hbar}
\end{equation}
Now, letting $y=z\hbar^{1/4}$ then the phase is in standard form after redefining the coefficients as 
\begin{align}
    X &=\left(\frac{-2}{4\sigma_p\alpha^2\sqrt{t}}-\frac{2bi\sqrt{t}}{\alpha^2\hbar}\right)\sqrt\hbar\\
    Y &=\frac{(1+i)x}{\hbar^{3/4}\alpha \, t^{1/4}}
\end{align}
Then the wave packet is precisely the Pearcey integral
\begin{align}
    \Psi(x,t)&=At^{-1/4}\bar{P}(X,Y)\label{eq46}\\
    \bar{P}(X,Y)&=\int_{-\infty}^{+\infty}e^{i(z^4+Xz^2+Yz)}dz
\end{align}
where $t^{-1/4}$ comes from the change of variables made.
The above definition of the Pearcey integral is valid (that is, the integral converges) only for $0\leq \arg(X)\leq \pi $ and $Y\in \mathbb{R}$ thus we need to do analytic continuation to define the Pearcey integral for $(X,Y)\in \mathbb{C}^2$. To achieve this, we rotate the integration path by $\pi/8$ and then we have
\begin{align}
    \bar{P}(X,Y) &=2e^{i\pi/8}P(Xe^{-i\pi/4},Ye^{i\pi/8})\\
    P(X,Y) &=\frac{1}{2}\int^{+\infty}_{-\infty}e^{-t^4+iYt-Xt^2}dt
\end{align}
Now, the integral is defined over all complex values for $X,Y$. The study of asymptotic expansion of this integral is carried out in different regions of $\mathbb{C}^2$ and details can be found in Refs. \cite{pearcey_complex,analytical_formulasofpearceyintegral}. Here we are interested in the leading order term of the Pearcey integral when $t$ is large. Considering the case where $b\neq 0$ for large $t$ it is enough to look at the asymptotics of the Pearcey integral for large $|X|$ and the following formula works when $\arg(X)\leq 3\pi/4$  this condition will eventually be satisfied for a sufficiently large $t$ because as $t\rightarrow\infty$, $X$ approaches a pure imaginary number which will have argument $\pi/2$. Therefore, the asymptotic formula is 
\begin{eqnarray}
    P(x,y) = \frac{1}{2}\sqrt{\frac{\pi}{x}}\,e^{-y^2/4x} \times \hspace{100pt} \nonumber \\
      \left(\sum_{k=0}^{n-1}\frac{(-1)^k}{k!(4x)^{2k}}H_{4k}\left(\frac{y}{2\sqrt{x}}\right)
    +R_n(x,y)\right)
\end{eqnarray}
The leading order term is 
$\dfrac{1}{2} \sqrt{\dfrac{\pi}{x}} e^{-y^2/4x}$ and now Taylor expanding the exponential term tells us that the large order behavior is like $x^{-1/2}$ since $x$ for large $t$ is proportional to $t^{1/2}$ so the integral behaves as $t^{-1/4}$ for large $t$. Then from Eq.\ref{eq46} we find that the wave packet has a decay behavior.
\begin{equation}
    \Psi(x,t)\sim t^{-1/2}
\end{equation}
This agrees with our calculation via singularity theory.

Next, we look at the case where $b=0$, that is, at the point of the bifurcation curve. From Eq.44, it is seen that, as $t\rightarrow\infty$, $|X| \rightarrow 0$. We therefore need the asymptotic formula for the Pearcey integral for small $|X|$. This is given by \cite{analytical_formulasofpearceyintegral},
\begin{equation}
P(x,y) = \sum_{k=0}^{n-1} P_k(y) \frac{(-x)^k}{k!} + R_n(x,y),
\end{equation}
Where $R_n(x,y)$ is the reminder term. An additional feature of this asymptotic expansion is that it is convergent, so the reminder term approaches zero as $n\rightarrow\infty$ for small $|x|$. The functions $P_k(y)$ are related to the generalized hypergeometric functions. We will list the first four of these, and the rest can be found from a recurrence relation.

\begin{widetext}
\begin{eqnarray}
P_0(y) &=& \frac{1}{4}\,\Gamma\left(\frac{1}{4}\right) {}_{0}F_{2}\left(-; \frac{1}{2}, \frac{3}{4}; \frac{y^4}{256}\right) - \frac{y^2}{8}\,\Gamma\left(\frac{3}{4}\right) {}_{0}F_{2}\left(-; \frac{5}{4}, \frac{3}{2}; \frac{y^4}{256}\right), \\
P_1(y) &=& \frac{1}{4}\Gamma\left(\frac{3}{4}\right) {}_{0}F_{2}\left(-; \frac{1}{2}, \frac{1}{4}; \frac{y^4}{256}\right) - \frac{y^2}{8}\Gamma\left(\frac{5}{4}\right) {}_{0}F_{2}\left(-; \frac{3}{4}, \frac{3}{2}; \frac{y^4}{256}\right), \\
P_2(y) &=& \frac{1}{4}\Gamma\left(\frac{5}{4}\right) {}_{1}F_{3}\left(\frac{5}{4}; \frac{1}{4}, \frac{1}{2}, \frac{3}{4}; \frac{y^4}{256}\right) - \frac{y^2}{8}\Gamma\left(\frac{7}{4}\right) {}_{1}F_{3}\left(\frac{7}{4}; \frac{3}{4}, \frac{5}{4}, \frac{3}{2}; \frac{y^4}{256}\right), \\
P_3(y) &=& \frac{1}{4}\Gamma\left(\frac{7}{4}\right) {}_{1}F_{3}\left(\frac{7}{4}; \frac{1}{4}, \frac{1}{2}, \frac{3}{4}; \frac{y^4}{256}\right) - \frac{y^2}{8}\Gamma\left(\frac{9}{4}\right) {}_{1}F_{3}\left(\frac{9}{4}; \frac{3}{4}, \frac{5}{4}, \frac{3}{2}; \frac{y^4}{256}\right).
\end{eqnarray}
\end{widetext}

The recurrence relation is given by
\begin{align}
    P_{k+4}(y)=&-\left(\frac{y}{4}\right)^2P_{k+1}(y)-\frac{(k+1)(2k+1)}{8}\,P_k(y) \notag\\&+
    \frac{(k+1)(4k+7)}{4}\,P_{k+2}(y),\ \   k=0,1,2,3,4...
\end{align}
The leading order term in this asymptotic expansion is thus $P_0(y)$ which has terms involving the generalized hypergeometric function ${}_0F_2(-;b_1,b_2;z)$ \cite{andrews1999special} defined as 
\begin{equation}
    {}_0F_2(-; b_1, b_2; z) = \sum_{n=0}^{\infty} \frac{1}{(b_1)_n (b_2)_n} \frac{z^n}{n!}
\end{equation}
since the expansion starts from $n=0$ the leading term of $P_0(y)$ is a constant $\frac{1}{4}\Gamma(\frac{1}{4})$. Then it follows from Eq.~\ref{eq46} that the wave packet has a decay behavior with a leading order term as
\begin{equation}
    \Psi(x,t)\sim t^{-1/4}.
\end{equation}
This again agrees with the calculation from singularity theory.
\section{Conclusion}\label{sec9}
Through our application of singularity theory to study wave packets in the $\kappa$-Minkowski space, we have established the following. In a $\kappa$-Minkowski space that has a deformed dispersion relation $H(p)=\gamma p^2+\alpha^4p^4/4=E$, where $\gamma=b\alpha^2+1/2m_{efff}$ the decay behavior of the wave packet undergoes a discontinuous change as $\gamma$ crosses a point of the bifurcation curve $\gamma=0$. The wave packet exhibits an anomalous decay behavior at the bifurcation point, as it changes the decay rate from $t^{-1/2}$ to $t^{-1/4}$.

This behavior can only be seen in a $\kappa$-Minkowski space time because the $p$ dependent modification we had introduced in $V(x)$ is unique to this space time. This result acts as a definite signature of noncommutative space time because the behavior is not arising from perturbative corrections to some standard physical quantities in the order of $\alpha$, but rather from a discontinuous change in the decay behavior of the wave packet. The question then is why this anomalous decay behavior is not observed. This is because of the difficulty of reaching the $\gamma=0$ value where it shows an anomalous decay. Note that $\gamma$ is not the tuning parameter of the system, it is $b$ which is related to $\gamma$ by $\gamma=b\alpha^2+1/2m_{efff}$. Thus, reaching $\gamma=0$ corresponds to $b$ being a large negative number (depending on how small $\alpha$ is). It turns out that for generic systems, $b$ may depend on energy and, as such, reaching a large negative value for $b$ will not be possible at the current available energy scales of experiments. In another of our works, we have shown that this energy scale depends inversely on $\alpha$. This answers the question of why such anomalous decay behavior is not seen in standard energy scales. This is also expected since noncommutativity of spacetime starts to matter only at energy scales where quantum gravity effects have to be considered.
\section{Acknowledgements}
We acknowledge the support of the NIUS programme of HBCSE-TIFR funded by the Department of Atomic
Energy, Govt. of India (Project No. RTI4001)
\bibliographystyle{apsrev4-2}
\bibliography{main}



\



\end{document}